\documentclass{ws-procs975x65}

\begin{document}

\title{THE PARADOX OF SOFT SINGULARITY CROSSING AVOIDED BY DISTRIBUTIONAL
COSMOLOGICAL QUANTITIES.}
\author{ALEXANDER KAMENSHCHIK}

\address{Dipartimento di Fisica e Astronomia and INFN, Via Irnerio 46, 40126 Bologna,
Italy\\
L.D. Landau Institute for Theoretical Physics of the Russian
Academy of Sciences, Kosygin str. 2, 119334 Moscow, Russia\\
E-mail: kamenshchik@bo.infn.it}

\author{ZOLT\'{A}N KERESZTES}
\address{Department of Theoretical Physics, University of Szeged, Tisza Lajos
krt 84-86, Szeged 6720, Hungary\\
Department of Experimental Physics, University of Szeged, D\'{o}m T%
\'{e}r 9, Szeged 6720, Hungary\\
E-mail: zkeresztes@titan.physx.u-szeged.hu}

\author{L\'{A}SZL\'{O} \'{A}. GERGELY}
\address{Department of Theoretical Physics, University of Szeged, Tisza Lajos
krt 84-86, Szeged 6720, Hungary\\
Department of Experimental Physics, University of Szeged, D\'{o}m T%
\'{e}r 9, Szeged 6720, Hungary\\
E-mail: gergely@physx.u-szeged.hu}

\begin{abstract}
A flat Friedmann universe filled with a mixture of anti-Chaplygin gas and
dust-like matter evolves into a future soft singularity, where despite
infinite tidal forces the geodesics can be continued. In the singularity the
pressure of the anti-Chaplygin gas diverges, while its energy density is
zero. The dust energy density however does not vanish, neither does the
Hubble parameter, which implies further expansion, if its evolution is to be
continuous. If so, the energy density and the pressure of the anti-Chaplygin
gas would become ill-defined, hence only a contraction would be allowed.
Paradoxically, the universe in this cosmological model would have to expand
and contract simultaneously. The paradox can be avoided by redefining the
anti-Chaplygin gas in a distributional sense. Then the Hubble parameter
could be mirrored to have a jump at the singularity, allowing for a
subsequent contraction. With this modification the set of Friedmann,
Raychaudhuri and continuity equations are all obeyed both at the singularity
and in its vicinity.
\end{abstract}





\bodymatter

\section{Introduction}

The discovery of the cosmic acceleration stimulated the development of
\textquotedblleft exotic\textquotedblright\ cosmological models of dark
energy; some of these models possess the so called soft or sudden
singularities characterized by the finite value of the radius of the
universe and of its Hubble parameter. One of examples of such singularities
is the Big Brake singularity arising in a specific tachyon model \cite{tach}%
. The toy tachyon model \cite{tach}, proposed in 2004, has two particular
features: $i)$ the tachyon field transforms into a pseudo-tachyon field; $%
ii) $ the evolution of the universe can encounter a new type of singularity
- the Big Brake singularity. When a universe encounter the Big Brake
singularity its scalefactor is finite, the velocity of expansion is equal to
zero, the deceleration is infinite. The predictions of the model match
observational data on supenovae of the type Ia \cite{tach1,tach2} and the
Big Brake singularity is a special one - it is possible to cross it \cite%
{tach2}.

One of the simplest cosmological models revealing the Big Brake singularity
is based on the anti-Chaplygin gas \cite{tach} with an equation of state 
\[
p=\frac{A}{\rho },\ \ A>0, 
\]%
which through the continuity equation leads to 
\[
\rho (a)=\sqrt{\frac{B}{a^{6}}-A},~B>0. 
\]%
At $a=a_{\ast }=\left( \frac{B}{A}\right) ^{1/6}$ the universe encounters
the Big Brake singularity. This singularity is traversable, because all the
Christoffel symbols are finite and hence the geodesics equations are
well-defined.

\section{The model with the anti-Chaplygin gas and dust}

Let us consider now the model of the flat Friedmann universe filled with the
anti-Chaplygin gas and dust \cite{paradox}. The energy density and the
pressure are 
\[
\rho (a)=\sqrt{\frac{B}{a^{6}}-A}+\frac{M}{a^{3}},\ \ p(a)=\frac{A}{\sqrt{%
\frac{B}{a^{6}}-A}}. 
\]%
Due to the dust component, the Hubble parameter has a non-zero value at the
singularity, therefore the presence of the dust implies further expansion.
With continued expansion however, the energy density and the pressure of the
anti-Chaplygin gas would become ill-defined.

We solve the paradox by redefining the anti-Chaplygin gas in a
distributional sense. Then a contraction could follow the expansion phase at
the singularity at the price of a jump in the Hubble parameter. Although
such an abrupt change is not common in any cosmological evolution, we show
that the set of Friedmann, Raychaudhuri and continuity equations are all
obeyed both at the singularity and in its vicinity.

The jump in the Hubble parameter 
\[
H\rightarrow -H 
\]%
leaves intact the first Friedmann equation $H^{2}=\rho $, the continuity
equations and the equations of state, however, it breaks the validity of the
second Friedmann (Raychaudhuri) equation $\dot{H}=-\frac{3}{2}(\rho +p)$.
This is, because in the vicinity of the singularity the Hubble parameter can
be expanded as 
\begin{eqnarray*}
H(t) &=&H_{S}sgn(t_{S}-t) \\
&&+\sqrt{\frac{3A}{2H_{S}a_{S}^{4}}}sgn(t_{S}-t)\sqrt{|t_{S}-t|}~,
\end{eqnarray*}%
leading to%
\[
\dot{H}=-2H_{S}\delta (t_{S}-t)-\sqrt{\frac{3A}{8H_{S}a_{S}^{4}}}\frac{%
sgn(t_{S}-t)}{\sqrt{|t_{S}-t|}}~. 
\]

To restore the validity of the Raychaudhuri equation we add a singular $%
\delta $ -term to the pressure of the anti-Chaplygin gas 
\[
p=\sqrt{\frac{A}{6H_{S}|t_{S}-t|}}+\frac{4}{3}H_{S}\delta (t_{S}-t). \label%
{pressure-new} 
\]
To preserve the equation of state we also modify the expression for its
energy density: 
\[
\rho=\frac{A}{\sqrt{\frac{A}{6H_{S}|t_{S}-t|}}+\frac{4}{3}H_{S}\delta
(t_{S}-t)}~. \label{en-den-new} 
\]

In order to prove that $p$ and $\rho $ represent a self-consistent solution
of the system of cosmological equations, we employed the following
distributional identities: 
\begin{eqnarray*}
\left[ sgn\left( \tau \right) g\left( \left\vert \tau \right\vert \right) %
\right] \delta \left( \tau \right) &=&0\ , \\
\left[ f\left( \tau \right) +C\delta \left( \tau \right) \right] ^{-1}
&=&f^{-1}\left( \tau \right) \ , \\
\frac{d}{d\tau }\left[ f\left( \tau \right) +C\delta \left( \tau \right) %
\right] ^{-1} &=&\frac{d}{d\tau }f^{-1}\left( \tau \right) \ .
\end{eqnarray*}

\section{Conclusion}

The use of generalized functions (distributions) is not uncommon in physics.
They appear naturally whenever there are lower-dimensional localized sources
(including branes in higher dimensional theories), but also when in quantum
field theory the product of distributions becomes well-defined by a
renormalization procedure. The addition of a $\delta $-function centred on a
point where the pressure already diverges can be considered as a similar
procedure.

Finally, we formulate questions for further related studies: How general is
the paradox of the soft singularity crossing? Is it possible to find other
ways out from the paradox of the soft singularity crossing?

This work was supported by  the
European Union / European Social Fund grant 
T\'{A}MOP-4.2.2.A-11/1/KONV-2012-0060 (ZK and L\'{A}G), and OTKA grant no. 100216
(ZK).

\end{document}